# Avoiding Intellectual Stagnation: The Starship as an Expander of Minds

I. A. Crawford

Department of Earth and Planetary Sciences, Birkbeck College London, Malet Street, London, WC1E 7HX

## Abstract

Interstellar exploration will advance human knowledge and culture in multiple ways. Scientifically, it will advance our understanding of the interstellar medium, stellar astrophysics, planetary science and astrobiology. In addition, significant societal and cultural benefits will result from a programme of interstellar exploration and colonisation. Most important will be the cultural stimuli resulting from expanding the horizons of human experience, and increased opportunities for the spread and diversification of life and culture through the Galaxy. Ultimately, a programme of interstellar exploration may be the only way for human (and post-human) societies to avoid the intellectual stagnation predicted for the 'end of history'.

## 1. Introduction

Twenty years ago the American political philosopher Francis Fukuyama [1,2] argued that the world is on a course of increasing political and cultural homogenization. Although events since Fukuyama first published his arguments indicate that the unification process, at least in its political dimension, is proceeding more slowly than he perhaps envisaged, a general trend towards political and cultural integration has been apparent in world history for several centuries and seems likely to continue. In many ways Fukuyama's argument is an essentially optimistic one, because a culturally and politically unified world would have fewer motivations for conflict and is therefore likely to be a more peaceful place.

Fukuyama himself was ambivalent towards this outcome, which he famously described as 'the end of history', because he believed that an end to competition between human societies would also mean an end to human achievement and creativity. As he put it [1]:

> "The end of history will be a very sad time. The struggle for recognition, the willingness to risk one's life for a purely abstract goal, the worldwide ideological struggle that called forth daring, courage, imagination, and idealism, will be replaced by economic calculation, the endless solving of technical problems, environmental concerns, and the satisfaction of sophisticated consumer demands. In the post-historical period there will be neither art nor philosophy, just the perpetual caretaking of the museum of human history."

In contrast to this rather depressing vision of the future, what we really want to do is build a human civilisation that is both stable and *dynamic*. That is, a civilisation which is at peace with itself, but which is nevertheless an exciting place in which to live and, most importantly, one whose history remains *open*. As I argued an initial response to Fukuyama's ideas [3], an ambitious programme of space exploration is ideally, and perhaps uniquely, suited to satisfying these socially desirable objectives. Humanity will very likely begin to experience these intellectual and cultural benefits by exploring and colonizing our own Solar System, but it is on the larger stage of interstellar exploration that they will really come into play.

In what follows I divide what I see as the potential interstellar sources of cultural and intellectual stimuli into the three headings of 'science', 'art' and 'philosophy', but I am aware that these distinctions are somewhat arbitrary – the cultural world, like the natural world, in reality forms a continuum of experience.

## 2. Science

There can be little doubt that science, especially in the fields of astronomy, planetary science and astrobiology, will be a major beneficiary of the development of an interstellar spaceflight capability [4,5]. In its long history, astronomy has made tremendous advances through studying the light that reaches us from the cosmos, but there is a limit to the amount of information that can be squeezed out of the analysis of starlight and other cosmic radiation. In many areas further progress will require *in situ* measurements of distant astronomical objects, which will only become possible in the context of a programme of interstellar space exploration. I have reviewed the scientific benefits of interstellar exploration in detail elsewhere [5], and so they are only briefly reiterate here. Broadly, they may be divided into the following sub-categories:

(1) Scientific investigations conducted on route (e.g. of the interstellar medium), and physical and astrophysical studies which could make use of interstellar spacecraft as observing platforms;

(2) Astrophysical studies of a wide range of different types of stars and their circumstellar environments;

(3) Planetary science studies of planets orbiting these stars, including moons, asteroids, and other small bodies;

(4) Astrobiological/exobiological studies of habitable (or inhabited) planets which may be found orbiting other stars.

A programme of interstellar exploration would enable scientific investigations in all of these areas, and could not fail to yield scientific knowledge unobtainable in any other way. This is especially true of planetary science and astrobiology because, although astronomical instruments based in the Solar System will become increasingly powerful and sophisticated over the coming centuries, and will certainly be able to reveal far more about the astronomical and astrophysical characteristics of stars and planets than is obtainable at present, many investigations simply cannot be pursued by remote sensing alone. Indeed, the history of the exploration of our own Solar System reveals that *in situ* measurements by spacecraft are required for the detailed geological, geophysical, and biological study of planets, and it seems clear that we will eventually require spacecraft to make *in situ* studies of other planetary systems as well.

The desirability of such direct investigations is especially apparent in the context of searching for life elsewhere in the universe. Cockell [6] has recently drawn attention to the likelihood that, given our lack of geological and geochemical knowledge of potentially habitable planets around other stars, even determining whether or not life is present on them may be impossible using astronomical means alone. Such an approach would be restricted to spectroscopic analyses of atmospheric gasses, with a view to identifying biology through the presence of trace biomarker molecules out of chemical equilibrium, but Cockell observes that:

> "For many atmospheric gases, the lack of knowledge about an exoplanet, including plate tectonics, hydrosphere-geosphere interactions, crustal geochemical cycling and gaseous sources and sinks, makes it impossible to distinguish a putative biotic contribution to the mixing ratios, regardless of the resolving power of the telescope" [6].

It follows that, for the vast majority of planets in the Galaxy, even determining if life is present or not will require *in situ* investigations. Moreover, even in those cases where Solar System-based astronomical observations of exoplanets are able to reveal putative biosignatures, definitive proof of life and follow-up studies of its underlying biochemistry, cellular structure, ecological diversity and evolutionary history will almost certainly require *in situ* measurements to be made. This in turn will require the transport of sophisticated scientific instruments across interstellar space.

The long-term societal and cultural consequences of these scientific discoveries are of course unpredictable, and will in any case depend on what actual discoveries are made. However, by opening up an unexplored, and potentially infinitely large, scientific frontier, from which a stream of new discoveries will constantly flow, a programme of interstellar exploration can hardly fail to keep scientific 'history' open and avoid any risk of intellectual stagnation as far as science is concerned.

## 3. Art

Without claiming to be an expert on the subject, I here take 'art' to include the creative activities of literature, poetry, painting, drawing, sculpture, music, dance, drama, computer generated imagery, etc, and perhaps also architecture, employed to convey subjective human values, experiences and emotions from one human (or indeed non-human or post-human) consciousness to others. In addition, there is an argument that art may also have an important role in mediating communication between emotional and rational levels *within* individual conscious minds; as McLaughlin [7] puts it "[t]he fine arts are modes of communication between our centres of thought and emotion and serve to assist in harmonizing these centres that must co-exist in one organism."

In terms of Karl Popper's 'three world' epistemology [8] we can therefore view art as facilitating communication of subjective information within and between conscious minds (which by Popper's definition inhabit 'World 2') through the manipulation of 'World 1' objects (i.e. the physical universe) to create 'World 3' products (i.e. consciously designed artefacts of various kinds) which in turn influence subjective experience in 'World 2'. These feedback loops between different levels of existence provide the conditions for the continued growth of both knowledge and art, and the continual enrichment of subjective experience [8].

McLaughlin [7] has given careful consideration to the potential impact of space exploration on the fine arts, and argued that that the influence is likely to be strong. Space exploration, and especially interstellar exploration, has the potential to stimulate the feedback loops between the different levels of experience identified by Popper by injecting new observations of the physical world ('World 1') and new subjective experiences ('World 2'). Many of these new artistic influences will manifest themselves as we explore our own Solar System [7], and some are already apparent, but they are likely to be even more profound on an interstellar scale.

At one level it seems obvious that new space scenes, and novel space events, must inspire new works of space art. It is difficult to see how this could be otherwise. However, the potential long-term impact of space, and especially interstellar, exploration is more profound. As I have noted previously (reference [3]; see also McLaughlin [7]), the increasing dominance of the "cosmic perspective" on human (and eventually post-human) thought is likely to change the whole paradigm of artistic expression. Not only will it be necessary to find ways of portraying and communicating human (and human-derived) values in the face of a universe whose vastness and strangeness will become ever more apparent as exploration proceeds, but the human (and post-human) mind is itself likely to become increasingly "cosmicized" [9] and this can hardly fail to be reflected in artistic evolution. Indeed, the American scholar Joseph Campbell grasped the power of space exploration to introduce a cosmic perspective into human affairs during the Apollo programme [10], and the consequences for our world view, and the artistic expression of it, can only become more profound as humanity moves out into the universe:

> "All the old bindings are broken. Cosmological centers now are any- and everywhere ... all poetry now is archaic that fails to match the wonder of this view" [10; p. 236].

**4. Philosophy**

The exploration of an open frontier, expanding into an effectively infinite universe, can hardly fail to stimulate philosophical investigations on multiple levels. Following my earlier discussion [3], I here make a distinction between moral and political philosophy, and give some examples of the aspects of both that I would expect to be stimulated by interstellar exploration. However, as I also pointed out in my earlier treatment [3], we must expect that this vast and mysterious universe very likely contains within itself the seeds of entirely new fields of philosophical investigation waiting to be discovered.

**4.1 Moral Philosophy**

We can already identify a number of moral-philosophical issues which may arise as a result of humanity's expansion beyond the Solar System:

• The ethics of confining human beings (or other conscious intelligent entities) on board slow, cramped, and presumably multi-generational, space vehicles will need to be explored if this option is pursued as a means of interstellar exploration and colonisation [11].

• The subject of environmental ethics will have to be extended to cover the interaction of humanity, and human-derived influences, with the material (and possibly biological) contents of other planetary systems. See the discussion in [12] for background on the types of ethical questions that may arise.

• The moral relationship between humanity and extraterrestrial life, in all of its probable diversity, will need to be addressed. Note that the mere *possibility* of coming across life elsewhere in the universe will act as a stimulus for moral philosophy, regardless of whether or not such life is actually discovered.

• Similar ethical problems may pertain to relationships between human explorers and their increasingly sophisticated, and possibly intelligent, retinue of robots and computers. Just what would be the moral rights and responsibilities of an artificially-intelligent interstellar space probe?

• Profound questions relate to the morality of colonising planets which harbour indigenous forms of life, or which may do so in the future. For example, at what point in the Earth's past, if any, would it have become morally unacceptable for an intelligent extraterrestrial civilisation to have colonised our planet? Any such colonisation would probably have precluded our own evolution. Does this mean that the colonisation of certain types of planet is unethical, and, if so, what types of planet?

• On the other hand, does there exist a moral *duty* for life in one part of the universe to spread it to parts where it is absent? Should we actively spread terrestrial life as far and wide as possible, just in case there is no life anywhere else? Far from the colonisation of other planets being immoral, is it perhaps a moral necessity?

Before we get very far out into the Galaxy we will need to have given serious thought to questions of this kind, even if absolute answers elude us.

**4.2 Political Philosophy**

As I pointed out in an earlier discussion [3], essentially all political philosophy to-date has been concerned with the organisation of, at most, a single planet. Often it has concerned with only tiny portions of that planet, such as city-states and nation-states. It seems clear that new fields of political speculation open up once the possibility of many inhabited worlds is admitted. As examples, consider the following questions:

• What would be the political status of interstellar colonies? Here the distances are so vast that that we might expect any attempt at political unification to be hopeless (at least in the absence of faster-than-light communication). But is this necessarily so? Are interstellar political institutions possible in principle? If not, is anarchy on interstellar scales inevitable [13]?

• Consider the political evolution of individual, isolated, interstellar colonies (or of equally isolated human populations on board slow interstellar vessels such as world-ships [11]). They might all be established as liberal-democracies, but would they remain so? Is there a danger of political backsliding into dictatorship, or of initially unified planetary colonies disintegrating into a multitude of warring states? Are there steps that could be taken to prevent this (see e.g. [13]), or should colonial independence be sacrosanct? For a thoughtful discussion of some of these issues see Cockell [14,15].

• If human colonisers should encounter comparably advanced extraterrestrial societies, would political relations (or even political union) be possible between them? If so, would it be desirable? Just what limits would biological differences place on the resulting political institutions?

Again, philosophical questions like these, and doubtless many others not considered here, will naturally arise as humanity (and/or post-humanity) spreads out into the universe.

**5. Expanding opportunities for the diversification of culture**

In addition to helping to prevent the stagnation of culture, the expansion of humanity (and ultimately post-humanity) into interstellar space would open up opportunities for the *diversification* of culture, what John Stuart Mill termed "different experiments of living" [16; p.120]. This was also recognized as a potential benefit of space colonization by Olaf Stapledon [17], when he expressed the view that:

> "The goal for the solar system would seem to be that it should become an interplanetary community of very diverse worlds each inhabited by its appropriate race of intelligent beings, its characteristic "humanity"..... Through the pooling of this wealth of experience, through this 'commonwealth of worlds' new levels of mental and spiritual development should become possible, levels at present quite inconceivable to man."

Although, as here envisaged by Stapledon, opportunities for diversification of culture would result from colonizing the planets and moons of the Solar System, the same basic argument applies to interstellar colonisation, which will presumably offer many more possibilities.

As Stapledon himself realized (see discussion in [18]), the scope for human (and post-human) colonization and diversification throughout the Galaxy depends crucially on the presence or absence of other intelligent species. We do not yet know how common, or otherwise, extraterrestrial intelligence may be, but the so-called Fermi Paradox (i.e. the observation that the Earth has not itself been colonized by other technological civilisations [19-21]) suggests that other civilisations may be rare, or even non-existent. If our Galaxy, or at least our part of it, really is devoid of other intelligent civilizations, it follows that the future of intelligence in the Galaxy will depend on *us*. It may then be desirable for humanity (or post-humanity) to start moving out through the Galaxy colonising uninhabited planets because this would enhance the diversity and creative potential of intelligent life in the universe.

**6. Exploration and Lockean–Popperian Epistemology**

In an article dedicated to philosophical issues related to interstellar space exploration, it seems appropriate to conclude with some philosophical observations of my own. In Section 3 I introduced Karl Popper's 'three world' epistemology [8] in the context of art as facilitator of the communication of subjective experience within and between conscious minds. However, Popper's epistemology has applications well beyond art theory. The basic feed-back loops he identified between the physical universe (World 1), subjective experience (World 2) and the objective constructs of conscious minds (World 3) apply to the whole body of human activity and knowledge. As Popper himself puts it:

> "There is also a most important feed-back effect from our creations upon ourselves; from the third world upon the second world. For the new emergent problems stimulate us to new creations …. [T]he feed-back of the third world upon the second, and even the first, are among the most important facts of the growth of knowledge" [8; p. 119].

However, it seems clear that this mechanism for generating knowledge, enhancing understanding, and enriching subjective experience, powerful though it is, would grind to a halt without the continual injection of new observations of the *real* world (i.e. from Popper's World 1).

Ultimately, all our science, art, and philosophy is built on what John Locke [22] called 'simple ideas' – that is ideas based on sense perception of the real world (World 1) and reflection on these perceptions (in World 2). Locke puts it thus:

> "All those sublime thoughts, which tower above the clouds, and reach as high as heaven itself, take their rise and footing here: In all that great extent wherein the mind wanders…. it stirs not one jot beyond those ideas which sense or reflection have offered for its contemplation" [22; p. 118].

In Locke's epistemology it follows that we cannot imagine genuinely new things, but instead have to *discover* them. This is what keeps Popper's feed-back loops running, and it is the ultimate cultural benefit of all exploration. Space exploration presents a vast new field of activity with literally infinite potential for discovery and intellectual stimuli of multiple kinds – certainly a far richer range of stimuli than we could ever hope to experience by remaining on our home planet. Humanity will begin to experience these intellectual and cultural benefits by exploring and colonizing our own Solar System, but it is on the far larger stage of interstellar exploration that they will really come into play.

## 7. Conclusion

Interstellar exploration and colonisation will advance human knowledge and culture in multiple ways. In particular, I have argued that a wide range of benefits will accrue to the broad cultural categories of science, art, and philosophy, and that these will help humanity, and our evolutionary successors, to avoid intellectual and cultural stagnation. Of course, all this does depend on interstellar travel being both physically possible and technically and economically practical for a sufficiently advanced technological civilisation. There are good reasons for believing that this will be the case (see references [23-27] for reviews), but not everyone who has considered this question accepts that humanity will ever be ready to make the leap across interstellar space. For example, Robinson [28] has argued that:

> "The stars exist beyond human time, beyond human reach. We live in the little pearl of warmth surrounding our star; outside it lies a vastness beyond comprehension. The solar system is our one and only home" [28; p. 328].

It is to be hoped that this will not turn out to be an accurate prophesy of humanity's future, because even a Solar System-wide society, such as Robinson depicts in his book, will eventually suffer the fate of the 'end of history' as envisaged by Fukuyama [1,2]. Although vastly larger and richer in intellectual and cultural stimuli than is Planet Earth alone, even a Solar System-wide civilisation will eventually prove to be an all-too-finite system. Only by eventually building starships, and exploring the "vastness beyond comprehension", will humanity (and post-humanity) permanently avoid the intellectual and cultural stagnation predicted for the 'end of history'.